\documentclass[10pt]{article}
\usepackage{amsfonts,amssymb,amsmath,amsthm}
\usepackage{graphicx}
\evensidemargin=\oddsidemargin \topmargin=0pt \advance\topmargin
by -\headheight \advance\topmargin by -\headsep

\newcommand{\dis}{\displaystyle}

\begin{document}

\thispagestyle{empty} \quad

\vspace{1cm}
\begin{center}

{\Large \textbf{On space-time noncommutative U(1) model at high
temperature\footnote{\small Talk given at the $8$th Workshop
"Quantum Field Theory Under the Influence of External Conditions",
 Leipzig, Germany, 17-21 September 2007.}}}

\vspace{1cm}

{\large Alexei Strelchenko}\\
{\large Dnepropetrovsk National University,\\
 49050
Dnepropetrovsk, Ukraine\\E-mail: alexstrelch@yahoo.com}

\vspace{1cm}

\end{center}

\begin{quote}
{\textbf{Abstract}}: {\small
 We extend the results of Ref.
\cite{SV} to noncommutative gauge theories at finite temperature.
In particular, by making use of the background field method, we
analyze renormalization issues and the high-temperature
asymptotics  of the one-loop Euclidean free energy of the
noncommutative $U(1)$ gauge model within imaginary time formalism.
As a by-product, the heat trace of the non-minimal photon kinetic
operator on noncommutative $S^1 \times R^3$ manifold taken in an
arbitrary background gauge is investigated. All possible types of
noncommutativity on $S^1 \times R^3$ are considered. It is
demonstrated that the non-planar sector of the model does not
contribute to the free energy of the system at high temperature.
The obtained results are discussed.}
\end{quote}
\vspace{1cm}


\section{Introduction}
Understanding fundamental properties  of  hot plasma in
noncommutative gauge theories, especially in NC QED, remains one
of the most challenging
 problems in  high-energy physics. Indeed, because of the noncommutative nature of
space-time,  even the simplest thermal $U(1)$ model exhibits such
odd  features as generation of the magnetic mass (associated with
noncommutative transverse modes), appearance of a tachyon in the
spectrum of quasi-particle excitations etc.
\cite{Arcioni:1999hw,Arcioni:1999gw,Fischler:2000fv,Fischler:2000fw,Landsteiner:2000bw,Landsteiner:2000bx,Brandt:2006,Brandt:2007}.
These observations concern mainly space/space noncommutative
theories where there are no notorious difficulties with causality
and unitarity \cite{unicas, GoMe}. At the same time, it was
realized that a space/space NC QFT may have non-renormalizable
divergences as a consequence of UV/IR mixing phenomenon
\cite{Gayral:2004cu} (see also \cite{Rivasseau} for recent
discussion).

The purpose of the present work is to gain some better insight
into basic aspects of the Euclidean-time formalism in thermal
gauge theories on NC $S^1\times R^3$, including renormalization
and the high-temperature asymptotic of the (Euclidean) free energy
(FE). For the sake of completeness, three different types of
noncommutative space-time will be worked out: namely, space/space,
full-rank and pure space/time noncommutativities. We begin our
analysis with the investigation of one-loop divergences in the
Euclidean NC $U(1)$ gauge model on $S^1\times R^3$  to make sure
that the theory does exist at least at the leading order of the
loop expansion. Then we will turn to the evaluation of the
high-temperature asymptotics of the one-loop FE. The main
attention will be paid to the non-planar sector of the
perturbative expansion. Thus, it was discovered in Ref.
\cite{Fischler:2000fv,Fischler:2000fw} that there is a drastic
reduction of the degrees of freedom in non-planar part of FE. Here
we will arrive at the same qualitative picture  for all types of
noncommutativity.

\section{The model}
 Consider $U(1)$ gauge model on NC $S^1\times {R}^3$.
Its  action  reads\footnote{As usual, we will work in the rest
frame of the heat bath with $u=(0,0,0,1)$, where $u$ is the heat
bath four velocity. All fields obey periodic boundary conditions.}
\begin{eqnarray}\label{1}
 S= -\frac{1}{4\,g^2} \int_{\mathcal M} d^4x~G_{\mu\nu}\star
 G_{\mu\nu},
\end{eqnarray}
where the integration is carried out over $\mathcal{M}=S^1\times
R^3$ manifold  and $G_{\mu\nu}$ denotes the curvature tensor of
$U(1)$ gauge connection.

To investigate quantum corrections to (\ref{1}) we employ the
background field method. To this aim we split the field $A_{\mu}$
into a classical background  field $B_{\mu}$
 and quantum
fluctuations $Q_{\mu}$, i.e.
 $A_{\mu}=B_{\mu} + Q_{\mu}$. Then,  substituting  this decomposition into (\ref{1}), we extract
 the part of the action (\ref{1}) that is quadratic in quantum
 fluctuations. In a covariant
background gauge it is written in the form (we use notations of
Ref. \cite{Strelchenko:2006za}):
\begin{equation}\label{3}
 S_{2}[B, Q,\overline{C}, C]= \int_{\mathcal M} d^4x~\left(-\frac{1}{2g^2}
Q_{\mu}(x)D^{(\xi)}_{\mu\nu}Q_{\nu}(x)+\overline{C}(x)~ D
C(x)\right),
\end{equation}
 where
\begin{equation}\label{4}
D^{\xi}_{\mu\nu}=-\Bigl[\delta_{\mu\nu}\nabla^{2}+(\frac{1}{\xi}-1)\nabla_{\mu}
 \nabla_{\nu} +2( L_{\Theta}(F_{\mu\nu})-R_{\Theta}(F_{\mu\nu}))\Bigr]
\end{equation}
is the photon kinetic operator and $D=-\nabla_{\mu}\nabla_{\mu}$
is the inverse propagator of ghost particles. Here $\nabla_{\mu}$
and $F_{\mu\nu}$ stand for the covariant derivative   and the
curvature tensor of  the background field
 $B_{\mu}$, respectively. Functional integration  of the
partition function  w.r.t. quantum fields gives the following
formal expression for the 1-loop effective action (EA),
\begin{eqnarray}\label{6}
  \Gamma^{(1)}[B]=\Gamma_{gauge}[B]+\Gamma_{ghost}[B]= \frac{1}{2}\ln \det\left(D^{\xi}\right)-
\ln\det(D).
\end{eqnarray}
 As  well-known this quantity is divergent and must be
regularized. This will be done by zeta-function regularization in
what follows.

For the study of thermal QFT one needs to introduce another
important object -- the free energy  of the system. Recall, that
there are two definitions of this quantity. One of them presents
the canonical FE,
\begin{equation}\label{CFE}
 { F^{C}(\beta)=\beta^{-1}\sum_{\omega}\ln\left(1-e^{-\beta
 \omega}\right)},
\end{equation}
which has clear physical meaning of "summation over modes". The
other one expresses FE in terms of the Euclidean EA,
\begin{equation}\label{EFE}
  {
   F^{E}(\beta)=\beta^{-1}\Gamma^{E}(\beta)},
\end{equation}
and is much more convenient from practical point of view.  These
two definitions are related by $$
   F^{E}(\beta)=F^{C}(\beta)+E_0,
$$ where ${E_0}$ is the energy of vacuum fluctuations.
 It should be noted, however, that a rigorous proof of this relation  even in
conventional field theories may be a highly non-trivial task (e.g.
for  thermal systems in curved spaces, see for instance Refs.
 \cite{nlsp,nlsp2}). The equivalence of the canonical and Euclidean FE  in QFT with space-time
noncommutativity (although with some heuristic  assumptions) was
discussed in Ref. \cite{SV}.

\section{Zeta-function regularization.}
 In the zeta regularization scheme, the regularized
EA (\ref{6}) is represented by
\cite{Elizalde:1994gf,Kirsten:2001ks,Vassilevich:2003xt}
\begin{eqnarray}\label{16a}
  \Gamma^{(1)}_{s}[B]= -\frac 12 \mu^{2s}
\Gamma(s)\left( \zeta \left(s,D^{\xi}\right)-2 \zeta (s,D)\right),
\end{eqnarray}
where $\zeta \left(s,D^{\xi}\right)$ and $\zeta \left(s,D\right)$
are zeta-functions of each operator in (\ref{6}), $s$ is a
renormalization parameter and $\mu$ is  introduced to render the
 mass dimension correct. The regularization is removed in the limit
$s\to 0$ giving
\begin{equation}
\Gamma^{(1)}_{s\rightarrow 0}[B]=-\frac 12 \left( \frac 1s
-\gamma_E +\ln \mu^2 \right)\zeta_{tot}(0) -\frac 12
\zeta'_{tot}(s),\label{17a}
\end{equation}
where $\gamma_E$ is the Euler constant and $\zeta_{tot}(s)=\zeta
\left(s,D^{\xi}\right)-2 \zeta (s,D)$.

To deal with the zeta-functions we need to introduce the heat
traces for the operators $D^{\xi}$ and $D$, respectively. Recall
that for a star-differential operator  $\mathcal{D}$ it is define
as
\begin{eqnarray}\label{14}
  K\left(t, \mathcal{D}\right)={\rm
  Tr}_{L^{2}}\left(\exp(-t\mathcal{D})-{\rm volume~ term}\right),
\end{eqnarray}
 where $t$ is the spectral (or "proper time") parameter. Symbol ${\rm Tr}_{L^2}$ denotes $L_2$-trace  taken on the space of square integrable functions ( on
$S^1 \times R^3$ with periodic boundary conditions in our case)
and may also involve  the trace over vector, spinor etc. indices.
The main technical result here is that on a (flat) NC manifold the
heat trace (\ref{14}) can be expanded in power series in small $t$
as:
\begin{equation}
K(t,\mathcal{D})=\sum_{n=1}^\infty t^{(n-4)/2} a_n(\mathcal{D}).
\,\label{18a}
\end{equation}
For further details, we refer the interested reader to Refs.
\cite{Vassilevich:2003yz,Gayral:2004ww,Gayral:2004cs,Vassilevich:2005yz,Gayral:2006vd,SV,Vassilevich:2007}.
Now, the zeta-function $\zeta_{tot}(s)$ has the following integral
representation,
\begin{eqnarray}\label{16b}
  \zeta_{tot}(s)=\frac{1}{\Gamma(s)} \int_0^\infty
  \frac{dt}{t^{1-s}}\left(
K^{\xi}\left(t, D^{\xi}\right)-2K\left(t, D\right)\right),
\end{eqnarray}
and to analyze the structure of (\ref{17a}) one should actually
evaluate the heat trace coefficients for each operator entering
(\ref{6}). For instance, taking into account  the relation $\dis
a_{k}(\mathcal{D})=\mathrm{Res}_{s=({4-k})/{2}}\Gamma(s)\zeta(s,
\mathcal{D}),$
 the
pole part of (\ref{17a}) can be re-expressed through the heat
trace coefficients  as
\begin{equation}\label{16}
\Gamma^{(1)}_{\rm pole}[B]=-\frac 12 \left( \frac 1s -\gamma_E
+\ln \mu^2 \right)\left( a_4(D^{\xi})-2a_4(D)\right).
\end{equation}
That is, on a 4-dimensional manifold it is determined by the 4th
heat trace coefficients.

\section{Evaluation of the heat trace coefficients}
To obtain the heat trace asymptotics of the non-minimal operator
(\ref{14})
       it is convenient to use the calculating method by Endo
\cite{Endo:1984sz} generalized on a NC case
\cite{Strelchenko:2006za}.
     Namely, if the background field satisfies the equation of
motion,
 the following relation holds\footnote{Notice that one has to eliminate volume divergences by
adding appropriate terms, cf. expr. (\ref{14}).}:
\begin{eqnarray}\label{19}
 K^{\xi}\left(t, D^{(\xi)}\right)=K^{\xi=1}\left(t, D^{(\xi=1)}\right)~~~~~~~~~~~~~~~~~~~~~~~~~~~~~~~~~~~~~~~~~~~~~~\\
 ~~~~~~~~~~~~~~-
\int_{t}^{\frac{t}{\xi}}d\tau~\int_{\mathcal M}
d^4x\,\left(\nabla_{\mu}\nabla'_{\mu}K(x, x';\tau|\beta)- {\rm
volume~term }\right)|_{x=x'},\nonumber
\end{eqnarray}
where $K(x, x';\tau|\beta)$ is the thermal heat operator of the
inverse ghost propagator.  Notice that RHS of this relation
consists of the heat traces of $minimal$ star-differential
operators. Calculating procedure for such objects is standard and
described, for instance, in Ref. \cite{Vassilevich:2005yz}. In
particular, it was found that the heat trace expansion for a
 generalized
star-Laplacian\footnote{That is, which includes both left and
right Moyal multiplications.}
 contains coefficients of two types:
 so-called planar and mixed heat trace coefficients. In our example, the first
 planar heat trace coefficient is given by
\begin{equation}\label{55}
  {a^{\rm pl. tot.}_4:=a_4(D^{(\xi)})-2a_4(D)=\frac{1}{16\pi^{2}}\left( -\frac{11}{3}\right)\int_{\mathcal{M}} d^4x \, F_{\mu\nu}\star
F_{\mu\nu}}.
\end{equation}
Evaluation of the mixed  heat trace coefficients, however, is more
involved. Here we inspect  three different cases.\\
  (i) Full-rank noncommutativity. To simplify computations   we assume that
the deformation matrix $\Theta$ has the canonical form:
\begin{eqnarray}\label{36}
\Theta= \left(
\begin{array}{cccc}
\theta S&0\\ 0&\vartheta S
\end{array}
\right),~~~~S= \left(
\begin{array}{cccc}
0&1\\ -1&0
\end{array}
\right).
\end{eqnarray}
However, the reader should be warned that, in general, a reference
frame where the matrix $\Theta$ has the block off-diagonal form
(\ref{36}) does not necessarily  coincide with the reference frame
of the heat bath. The first nontrivial mixed coefficient can be
now easily evaluated and has the form (see also \cite{SV} for some
technical details)
\begin{eqnarray}\label{40}
 { a^{\rm mix.tot.}_5=-\frac
{\xi^{-1/2}}{2\beta\,\theta^2\pi^{5/2}} \sum_{n\in {Z}}\int_{{R}^2
\times S^1} dx_{\perp}dx^4\int_{{R}^2 \times S^1}
dy_{\perp}dy^4\int_{R} dx^3\times}~~~~~\\ {~~~~~~~~\times
\sum_{\mu,\, \mu\neq 3}B_\mu\left(x^{1}, x^{2},
x^3+\frac{\pi|\vartheta|
  n}{\beta};x^4\right)\,
B_\mu\left(y^{1}, y^{2}, x^3-\frac{\pi|\vartheta|
  n}{\beta};y^4\right)}.\nonumber
\end{eqnarray}
This coefficient is divergent as $\theta \rightarrow 0$ and/or
$\vartheta \rightarrow 0$ that is a manifestation of the
well-known UV/IR phenomenon
\cite{Minwalla:1999we,Chepelev,Arefeva}. \\ (ii) Pure space/time
noncommutativity ($\Theta^{ij}=0$ and $\Theta^{i4}$ is  directed
along $x_{\|}$ axis). In this case
 the first mixed heat trace coefficient is presented by\\
\begin{eqnarray}\label{50}
{a^{\rm mix.tot. }_3=-\frac {1}{2\beta\,\pi^{3/2}}
\left(2-\sqrt{\xi}\right)\sum_{n\in {Z}}\int_{S^1 \times S^1}
dx^4dy^4 \int_{R^3}dx_{\perp}\, dx_\| ~\times}~~~~~~~~\\
~~~~~~~~~~~~~~~{\times B_4\left(x_{\perp},
x_\|+\frac{\pi|\vartheta|
  n}{\beta};x^4\right)\,
B_4\left(x_{\perp}, x_\|-\frac{\pi|\vartheta|
  n}{\beta};y^4\right)}.\nonumber
\end{eqnarray}
(iii) Space/space noncommutativity ($\Theta^{ij}\neq0$,
$\Theta^{4i}=0$). One finds
\begin{eqnarray}\label{54}
{ a^{\rm mix. tot.}_{4}=\frac{(\ln{\xi}-2)}{8\,
\theta^2\pi^{3}}\int_{S^1 \times {R}}dx^3dx^4~\int_{{R}^{2} \times
{R}^{2}}
 dx_{\perp}~dy_{\perp}\times~~~~~~~~~~~~~~~~~~~~} \nonumber\\~{\times\sum_{i=1,2}B_{i}(x_{\perp}, x^3; x^4) ~
 B_{i}(y_{\perp}, x^3; x^4)}.
\end{eqnarray}
 From
(\ref{16}) we see that this coefficient does contribute to the
pole term of the one-loop EA and, hence, affects renormalization
of the model that will be explained in a moment.

\section{Renormalization and high-temperature asymptotics}
Let us now look a little more closely at the divergent part of EA
(\ref{16}). Clearly,  in the case of noncommutative compact
dimension it is defined solely by the planar heat trace
coefficient (\ref{55}). That is, the pole part of the one-loop EA
has the form
\begin{equation}\label{55a}
  \Gamma^{(1)}_{\rm pole}[B]=-\frac 1{2s} \int_{\mathcal{M}} d^4x \left(-(4\pi)^{-2} \frac{22}{6}\, F_{\mu\nu}\star
F_{\mu\nu}\right),
\end{equation}
leading thus to  the standard renormalization group. We see that
the source of the UV divergence in (\ref{17a}) is associated with
the original four-dimensional field theory and this divergence is
removed by ordinary renormalization at zero temperature. However,
the situation changes drastically when the compact coordinate is
commutative: in this particular case the expression (\ref{55a})
contains an additional term due to the mixed heat trace
coefficient (\ref{54}). Although this new term is also temperature
independent, it brings into EA a non-local and, moreover,
gauge-fixing dependent divergence which cannot be eliminated by
any renormalization prescription.

To obtain high-temperature asymptotics of the one-loop EA we
rewrite (\ref{16a}) as
\begin{eqnarray}\label{58}
  \Gamma^{(1)}_{s}[B]= \mu^{2s} \sum_{k=2}\int_0^\infty
\frac{dt}{t^{3-s}}t^{\frac{k}{2}}\left( \left(-\frac 12
a_{k}(D^{\xi})+a_{k}(D)\right)+\right.~~~~~~~~~~~~~~~~~~~~~~~~~~~~~~~~~~~~~\nonumber\\
\qquad \left.+2\sum_{n=1}
 e^{-\frac{\beta^{2}n^{2}}{4t}}\left(-\frac 12
a^{\rm planar}_{k}(D^{\xi})+a^{\rm planar}_{k}(D)\right)\right) ,
\end{eqnarray}
where we retained all exponentially small terms in the planar
sector as well. (They must be taken into account  when the
parameter $\beta$ is small). The evaluation of the planar part
proceeds exactly as in the conventional thermal $SU(2)$
gluodynamics giving
\begin{eqnarray}\label{59}
{S_{\rm tree}[B]+ \Gamma^{(1)}_{\rm planar}[B]\simeq \left(-\frac
{1}{4g_{R}^2(T)}\int_{\mathcal{M}} d^4 x F_{\mu\nu}\star
  F_{\mu\nu}\,
 \right.}~~~~~~~~~~~~~~~~~~~~~~~~~~\\
 \qquad {\left.+\sum_{k=6}\left(\frac{\beta}{2} \right)^{2k-4}\left(a^{\rm planar}_k(D^{\xi})-2a^{\rm planar}_k(D)\right)\zeta(2k-4)\Gamma
 (k-2)\right)},\nonumber
\end{eqnarray}
from which one deduces high temperature behaviour of NC $U(1)$
effective coupling:
\begin{equation}\label{echarge}
 {g_{R}^2(T)=g_{R}^2\left(1+\frac{g_{R}^2}{4\pi^2}\frac{11}{3}\ln\left(T/T_0\right)\right)^{-1}}.
\end{equation}
 It should be emphasized, however, that the formula (\ref{echarge})
makes sense unless a compact dimension is commutative: as we have
already seen, within space/space NC $U(1)$ model one cannot
renormalize the charge because of the non-planar contribution
(\ref{54}).

Now consider the non-planar part of EA. For the sake of
definiteness let us focus on the pure space/time noncommutativity.
  First of all, we note
that the expression  (\ref{50})
 is valid whenever the
condition $\dis {|\vartheta|}/{\beta}\neq 0$ holds. Hence, it is
interesting to explore high temperature regime when $\dis
{|\vartheta|}/{\beta}\gg  C_0,~~ C_0 \in {R}_{+}$. We assumed that
the background field $B_{\mu} \in C^\infty (S^1\times {R}^3)$ and,
therefore, it should vanish exponentially fast at large distances.
For $n\neq 0$ one estimates $$B_\mu\left(x_{1}, x_{2},
z+\frac{\pi|\vartheta|
  n}{\beta};x_4\right)\,
B_\mu\left(y_{1}, y_{2}, z-\frac{\pi|\vartheta|
  n}{\beta};y_4\right)\sim C_2 \exp{\left(-C_1 \frac{|\vartheta|}{\beta}\right)}, ~~~ \frac{|\vartheta|}{\beta}\gg  C_0,$$
  where $C_1$ is a positive constant which characterizes the
fall-off of the gauge potential at large distances.  Up to an
inessential overall constant the contribution of the first mixed
coefficient to the effective potential can be estimated as
\begin{eqnarray}\label{60}
 a^{{\rm tot}}_3=\frac
{\left(1+\sqrt{\xi}\right)}{2\beta(\pi)^{3/2}}  \int_{S^1 \times
S^1} dx^4dy^4 \int_{{R}^3} dx\, B_4\left(\bar{x};x^4\right)\,
B_4\left(\bar{x};y^4\right).
\end{eqnarray}
 Notice that
this expression is insensitive to the value of the deformation
parameter\footnote{Of course, this does not mean that the
expression (\ref{6}) possesses a smooth commutative limit: in
obtaining high-temperature asymptotics for (\ref{60}) we assumed
$|\vartheta|\neq 0$.}. Moreover, since in the limit $\beta
\rightarrow 0$ the main contribution to (\ref{60}) comes from the
zero bosonic modes, the mixed heat trace coefficients behave as $
\sim \beta C$, where $C$ is some temperature-independent quantity.
From the definition (\ref{EFE}) it follows that, at least on the
one-loop level, the non-planar part of EA provides the
temperature-independent contribution to the Euclidean FE and
therefore can be neglected in the high temperature limit.

\section{Conclusion}
 In this paper we have
 investigated the one-loop quantum corrections to EA (resp. Euclidean FE) in NC thermal $U(1)$
theory  within the imaginary time
 formalism.   Let us summarize the obtained results.

 First, in
 the space/space noncommutative QED, the
 renormalizability of the theory is ruined by the non-planar
 sector of the perturbative expansion. This phenomenon was already
 observed, for instance, in Ref. \cite{Gayral:2004cu} (see also \cite{Gayral:2006vd,Strelchenko:2006za}).
 At the same time, in the case of a noncommutative compact dimension
   the theory can be renormalized, at least on one-loop level, by the standard  renormalization
   prescription.

   Second, we calculated the heat trace asymptotics for the non-minimal photon  kinetic
operator on NC $S^1\times {R}^3$. We saw, in particular, that the
noncommutativity of the compact coordinate results in arising of
additional  odd-numbered  coefficients in the heat trace
expansion. Furthermore, in the case of pure space/time
noncommutativity the first nontrivial mixed contribution to the
heat trace  appears in $a^{\rm mixed}_3$. Although this
coefficient does not affect counterterms in the zeta function
regularization, it can lead to certain troubles in different
regularization schemes, see Ref.\cite{SV} for further discussion.

Third, we obtain the high-temperature asymptotics of the one-loop
Euclidean FE (\ref{EFE}). It is rather remarkable that the
non-planar sector does not contribute at high temperature for any
type of noncommutativity. This seems to be in accordance with
observations made in earlier works where a drastic reduction of
the degrees of freedom in non-planar part of FE was discovered
\cite{Fischler:2000fv,Fischler:2000fw}. There is a subtlety,
however, that one should keep in mind. Namely, if noncommutativity
does not involve time, there are no difficulties in developing the
Hamiltonian formalism for  a NC theory and equivalence of the
canonical and Euclidean free energies is proved by standard
arguments \cite{DowKen}. Contrary to this, in the space/time NC
theories there is no good definition of the canonical Hamiltonian
and, consequently, of the canonical FE (\ref{CFE}) although some
progress in this direction has been made recently in
Ref.\cite{SV}.

Finally, an extension of our results to more general case of U(N)
gauge symmetry can be done straightforwardly. Indeed, one can show
that the mixed heat trace coefficients are completely determined
by U(1) part of the model \cite{Strelchenko:2006za}. In the
diagrammatic approach this implies the known fact that non-planar
one-loop U(N) diagrams contribute only to the U(1) part of the
theory \cite{Minwalla:1999we,Armoni:2000xi}.

\subsection*{Acknowledgments}

\hspace{\parindent}I am very grateful  to Michael Bordag for
inviting me and sponsoring my participation in Workshop QFEXT07,
Leipzig 2007. I wish to thank Dmitri Vassilevich for many
elucidative comments and discussions   on the ideas reported here.


\end{document}